\documentclass[
journal=jctcce,
manuscript=article,
layout=onecolumn,
linecount
]{achemso}

\usepackage{tocloft}
\usepackage{graphicx}  
\usepackage{dcolumn}   
\usepackage{bm}        
\usepackage{amssymb}   
\usepackage{amsmath}
\usepackage{float}
\usepackage{afterpage}
\usepackage{hyperref}
\usepackage{placeins}
\usepackage{chemformula}

\usepackage{mathtools}
\usepackage{graphicx}
\usepackage{dcolumn}
\usepackage{bm}
\usepackage{lipsum}
\usepackage[utf8]{inputenc}
\usepackage[T1]{fontenc}
\usepackage{mathptmx}
\usepackage{etoolbox}
\usepackage{orcidlink}
\usepackage{physics}

\usepackage{graphicx}
\usepackage{dcolumn}
\usepackage{bm}
\usepackage{subcaption}
\usepackage{multirow}
\usepackage{soul}
\usepackage[normalem]{ulem}
\usepackage{xcolor}
\usepackage{kantlipsum}
\usepackage[version=4]{mhchem}
\usepackage{textcomp}

\usepackage{booktabs}
\newcolumntype{C}{>{$}c<{$}}
\AtBeginDocument{
\heavyrulewidth=.08em
\lightrulewidth=.05em
\cmidrulewidth=.03em
\belowrulesep=.65ex
\belowbottomsep=0pt
\aboverulesep=.4ex
\abovetopsep=0pt
\cmidrulesep=\doublerulesep
\cmidrulekern=.5em
\defaultaddspace=.5em
}
\usepackage{siunitx}
\usepackage{multirow}

\usepackage{xfrac}
\usepackage{orcidlink}
\usepackage{xcolor}

\definecolor{amber}{rgb}{1,0.49,0}

\newcommand{\editor}[2]{%
  \expandafter\newcommand\csname #1note\endcsname[1]{%
    \textcolor{#2}{(\textbf{#1:} ##1)}}%
  \expandafter\newcommand\csname #1\endcsname[1]{%
    \textcolor{#2}{##1}}%
  \expandafter\newcommand\csname #1cancel\endcsname[1]{%
    \textcolor{#2}{\sout{##1}}}%
  \expandafter\newcommand\csname #1change\endcsname[2]{%
    \textcolor{#2}{\sout{##1} ##2}}%
  \newenvironment{#1text}{\color{#2}}{\color{black}}
}

\editor{ED}{purple}
\definecolor{verde}{rgb}{0.,0.6,0}
\editor{PP}{verde}
\editor{SB}{amber}
\editor{resub}{cyan}

\usepackage{mathrsfs}
\usepackage{placeins}

\graphicspath{ {./images/} }

\makeatletter
\patchcmd{\acs@contact@details}{E}{*\,E}{}{}
\makeatother

\author{Giacomo Tenti} 
\email{gtenti@sissa.it}
\affiliation{International School for Advanced Studies (SISSA),
Via Bonomea 265, 34136 Trieste, Italy}

\author{Kousuke Nakano}
\affiliation{Center for Basic Research on Materials, National Institute for Materials Science (NIMS), 1-2-1 Sengen, Tsukuba, Ibaraki 305-0047, Japan}
\alsoaffiliation{Center for Emergent Matter Science (CEMS), RIKEN, 2-1 Hirosawa, Wako, Saitama, 351-0198, Japan.}

\author{Michele Casula}
\affiliation{Institut de Minéralogie, de Physique des Matériaux et de Cosmochimie (IMPMC), Sorbonne Université, CNRS UMR 7590, MNHN, 4 Place Jussieu, 75252 Paris, France}
\date{}

\title{Self-consistency error correction for accurate machine learning potentials from variational Monte Carlo}
\begin{document}

\maketitle

\begin{abstract}

Variational Monte Carlo (VMC) can be used to train accurate machine learning interatomic potentials (MLIPs), enabling molecular dynamics (MD) simulations of complex materials on time scales and for system sizes previously unattainable.
VMC training sets are often based on partially optimized wave functions (WFs) to circumvent expensive energy optimizations of the whole set of WF parameters. However, frozen variational parameters lead to VMC forces and pressures not consistent with the underlying potential energy surface, a bias called the self-consistency error (SCE). Here, we demonstrate how the SCE can spoil the accuracy of MLIPs trained on these data, taking high-pressure hydrogen as test case. We then apply a recently introduced SCE correction [{\it Phys. Rev. B} \underline{109}, 205151 (2024)] to generate unbiased VMC training sets based on a Jastrow-correlated single determinant WF with frozen Kohn-Sham orbitals.
The MLIPs generated within this framework are significantly improved and can approach in quality those trained on datasets built with fully optimized WFs.
Our conclusions are further supported by MD simulations, which show how MLIPs trained on SCE-corrected datasets systematically yield more reliable physical observables. Our framework opens the possibility of constructing extended high-quality training sets with VMC.
\end{abstract}

\section{Introduction}
Since their introduction~\cite{Blank1995,Behler2007}, machine learning interatomic potentials (MLIPs) have revolutionized the field of molecular dynamics (MD)~\cite{Handley2010,Behler2011,Manzhos2015,Botu2017,Deringer2019,Wood2019,Dral2020,Noé2020,Behler2021,Unke2021}. 
These models can learn the potential energy surface (PES) of a given system starting from a representative set of configurations, called the training set. Each element of this set is labeled by its physical properties (e.g., energy, ionic forces, pressure, etc.), usually computed using \emph{ab initio} electronic structure methods. 
After it is trained, an MLIP is capable of reproducing the observables of the system on new configurations with the accuracy of the original method but with a computational cost of several orders of magnitude smaller~\cite{Behler2007}. 
This has opened the possibility of running MD simulations at \emph{ab initio} accuracy with sizes and lengths that were previously out of reach~\cite{Jia2020}. 
By far, the most common method used for generating MLIPs datasets is density functional theory (DFT)~\cite{Martin2004}, which can easily produce $\sim10^4$-$10^5$ training points with modest computational resources, usually sufficient to train data-heavy machine learning frameworks~\cite{Deringer2019,Schleder2019}. For several systems, however, DFT fails to capture the strong effect of electronic correlations or may provide results that strongly depend on the choice of the exchange-correlation functional. 
Quantum chemistry methods, such as the density matrix renormalization group (DMRG) \cite{Olivares-Amaya2015}, and the complete active space self-consistent field (CASSCF)\cite{Jensen1987,Werner1985}, provide a much better description of correlation, but their applicability to periodic extended systems is limited by their large computational cost. Because of this, even if significant progress has been made recently, also thanks to the implementation of new machine learning schemes \cite{Rath2025}, the range of applicability of these approaches remains limited to a few tens of atoms and a few $ps$-long trajectories.
Quantum Monte Carlo (QMC) methods are a series of stochastic algorithms that provide state-of-the-art results for a large class of systems, from molecules to solids~\cite{Foulkes2001,Becca2017}, while being less computationally demanding.
Despite this, QMC approaches still have a cost about 10-100 times larger than DFT. 
In the context of MLIPs, this implies that the number of training configurations that can be generated within a given amount of resources is generally smaller than the one attainable with mean-field methods. Recently proposed solutions to this problem make use of data-efficient schemes in combination with hierarchical machine learning (or $\Delta$-learning~\cite{Ramakrishnan2015}) to construct QMC-based MLIPs~\cite{Tirelli2022,Tenti2024}.
However, within the most popular QMC approach, i.e., variational Monte Carlo (VMC), another issue prevents a straightforward application of MLIPs, presented in what follows.

The efficiency of VMC is largely due to the complexity of the wave function (WF) employed in the calculations to describe the electronic part. A possible choice for the WF $\Psi$ is
\begin{equation*}
    \Psi = \exp \left( J \right)  \times \Phi_{\textrm{AS}},
\end{equation*}
where $\exp \left( J \right)$ is a bosonic term (i.e., symmetric under particle exchange) called the Jastrow factor, and $\Phi_{\textrm{AS}}$ is an antisymmetric (AS) part. 
Both terms contain variational parameters, whose number can lead to flexible WFs.
However, there is a practical difficulty in optimizing a large number of variables in a stochastic setting, particularly those belonging to the AS part.
Thus, only the Jastrow parameters are usually optimized, and $\Phi_{\textrm{AS}} \equiv \Phi_{\textrm{SD}}$ is often taken as a single Slater determinant (SD) obtained from a DFT calculation~\cite{Becca2017}. 
This gives rise to the frozen-orbital Jastrow Slater determinant (JSD) wave function, an appealing candidate for VMC-based machine learning applications, since it provides an excellent tradeoff between accuracy and computational cost. Indeed, the Jastrow optimization is much less demanding than the full optimization also involving the determinant.  
Nevertheless, the use of the frozen Slater determinant introduces a bias in the VMC forces and pressures computed with this WF.
This bias has been dubbed as the self-consistency error ~\cite{Tiihonen2021,Nakano2024},  and we are going to show that it is particularly relevant in the context of VMC-based MLIPs. 
Even though machine learning potentials are, by definition, consistent, a biased training set can nonetheless spoil the final accuracy of the model. A solution would be to use only the energies for the training; this, however, usually increases by orders of magnitude the number of configurations necessary to reach a given accuracy~{\cite{Chmiela2017, Christensen2020}}.

In Ref.~\citenum{Nakano2024}, a suitable correction was developed to remove the self-consistency error, adding a relatively small computational overhead to the calculation. 
The aim of this work is to show the actual impact of the self-consistency error on an MLIP in a working case scenario, and to demonstrate how its correction can be applied to generate unbiased VMC datasets with a JSD WF for machine-learning applications. 
As a test case, we consider high-pressure hydrogen, a widely studied system with applications ranging from planet modeling~\cite{Helled2020} to inertial confinement fusion~\cite{Hu2011,Craxton2015,Hu2015}, and often used to benchmark numerical methods~\cite{Bonitz2024}. It is also a system where DFT usually fails, when compared with the experimental data\cite{Monacelli2023}, making it an ideal testbed for our purposes. In particular, we focus on density and temperatures relevant for the Hugoniot curve, \emph{i.e.}, the set of possible states reachable with a shock wave~\cite{Duvall1977}. The deuterium Hugoniot was recently computed in Ref.~\citenum{Tenti2024} using MLIPs trained on accurate QMC data. 
There, the self-consistency error was avoided by relaxing the full set of variational parameters in the WF, \emph{i.e.}, including both the Jastrow factor and the antisymmetric part. The resulting optimization improves the consistency of the forces and pressures and also changes the underlying PES, at 
a computational cost
about 10 times larger than the standard frozen-orbital JSD optimization.

Here, we explore a different path, by showing that a very similar accuracy can be obtained by training models on a frozen-orbital JSD dataset supplemented by the self-consistency error correction. 
In particular, we  
compute VMC forces and pressures on the same configurations using a JSD WF, with and without corrections, and use them to train the different models. 
The performances of these MLIPs are
compared with the one of an MLIP trained on the fully optimized dataset of Ref.~\citenum{Tenti2024}, which we take as the reference. In this way, we show the relevance of the self-consistency error in the MLIP generation and the need for its correction, 
by studying 
the impact of both the correction and the 
full
optimization of the WF on thermodynamic equilibrium quantities such as pressure and 
radial distribution functions. 
This work demonstrates the 
importance
of obtaining unbiased QMC datasets for MLIPs at the VMC level using cheap-to-optimize JSD-type WFs, including also ionic forces and stress components.

\section{Methods}\label{sec:methods}

We briefly present the main methods used in this work. 
In Sec.~\ref{sec: generalities MLIP} we introduce 
general aspects of MLIPs, focusing on 
one
specific framework, 
\emph{i.e.,} the kernel ridge regression, and on the loss function used for model training. In Sec.\ref{sec: SCE correction}, we 
review the forces and pressure correction developed in Ref.~\citenum{Nakano2024} to solve the 
self-consistency error
in VMC and describe its implementation in our WF representation.

\subsection{Generalities of MLIPs}\label{sec: generalities MLIP}
The basic goal of any MLIPs is to predict the energy of a given $N$-atom configuration described by its ionic positions, i.e., $E = E \left( \mathbf{R}_1, \cdots , \mathbf{R}_N\right)$. We now suppose that $E$ can be expressed as a sum of atomic contributions~\cite{Behler2007}, each depending on the relative coordinates of all the other atoms with respect to the 
$i$-th
one: 
\begin{equation}
E = \sum_{i=1}^N e ( \mathcal{R}_i) \qquad \text{where} \quad \mathcal{R}_i = \left(\mathbf{R}_{i1} , \dots , \mathbf{R}_{iN} \right) \quad \text{with}\quad \mathbf{R}_{ij} \equiv \mathbf{R}_j - \mathbf{R}_i. \label{eq: energy decomposition}
\end{equation}
Here, we 
refer to $\mathcal{R}_i$ as the local environment around the $i$-th atom and to $e\left(\mathcal{R}_i\right)$ as the local atomic energy, and suppose the minimum image convention holds when periodic boundary conditions are applied.
Often, the function $e\left(\mathcal{R}_i\right)$ in Eq.\eqref{eq: energy decomposition} 
is considered local, i.e., only dependent on the atoms closer than a certain cutoff radius $r_c$.
Different types of MLIPs may be distinguished by their internal representation of the local atomic energy. Many techniques have been successfully applied over the years, including deep neural networks~\cite{Behler2007,Behler2011,Behler2021}, Gaussian approximation potentials~\cite{Bartok2010,Bartok2015}, kernel ridge regression~\cite{Rupp2012,Tirelli2022},   graph neural networks~\cite{ Batzner2022,Batatia2022}, and more. 
Independently of the specific architecture, $e\left(\mathcal{R}_i\right)$ 
usually depends on a set of model parameters $\mathbf{p}$ which have to be optimized: 
\begin{equation}
    e\left(\mathcal{R}_i\right) \equiv e\left(\mathcal{R}_i  ; \mathbf{p}\right).
\end{equation}
Most MLIPs implement a functional form that already satisfies some of the symmetries required by physical constraints, such as permutational and rotational invariance. 
To determine the model parameters $\mathbf{p}$, we consider a set of $N_{\textrm{train}}$ configurations, \emph{i.e.}, the training set, for which we compute the desired physical quantities using a target numerical method. 
In particular, we indicate with $\mathcal{R}^{\mu}_i \equiv (\mathbf{R}^{\mu}_1 -\mathbf{R}^{\mu}_i , \dots , \mathbf{R}^{\mu}_{N_{\mu}}-\mathbf{R}^{\mu}_i)$ the $i$-th atomic environment belonging to the $\mu$-th configuration of the set, with $\mu = 1, \dots , N_{\textrm{train}}$, having $N_{\mu}$ atoms and volume $V_{\mu}$ (for simplicity we will consider the supercell to be cubic). Moreover let $E^{\mu}_{\textrm{ref}}$, $\mathbf{F}_{j,\textrm{ref}}^{\mu}$ and $P^{\mu}_{\textrm{ref}}$ be the 
values of the total energy, the force acting on the $j$-th atom, and the (isotropic) virial pressure of the configuration. 
The model parameters $\mathbf{p}$ are then optimized by minimizing a loss function such as
\begin{align}
    \mathcal{L}(\mathbf{p}) & =   W_E \frac{1}{N_{\textrm{train}}} \sum_{\mu = 1}^{N_{\textrm{train}}} \left[ \frac{1}{N_{\mu}}\sum_{i= 1}^{N_{\mu}} e\left(\mathcal{R}_i^{\mu}; \mathbf{p}\right) - \frac{1}{N_{\mu}} E^{\mu}_{\textrm{ref}} \right]^2  \notag\\
    & + W_F \frac{1}{N_{\textrm{train}}} \sum_{\mu = 1}^{N_{\textrm{train}}} \frac{1}{3 N_{\mu}} \sum_{j = 1}^{N_{\mu}}\sum_{\alpha = x,y,z}\left[ - \frac{\partial}{\partial R^{\mu}_{j, \alpha}}\sum_{i = 1}^{N_{\mu}} e\left(\mathcal{R}_i^{\mu}; \mathbf{p}\right) - F^{\mu}_{j , \alpha, \textrm{ref}} \right]^2 \notag\\
     & +   W_P \frac{1}{N_{\textrm{train}}} \sum_{\mu = 1}^{N_{\textrm{train}}} \left[ - \frac{\partial}{\partial V_{\mu}}\sum_{i = 1}^{N_{\mu}} e\left(\mathcal{R}_i^{\mu}; \mathbf{p}\right) - P^{\mu}_{\textrm{ref}} \right]^2. \label{eq.5: loss function}
\end{align}

The three terms in Eq.\eqref{eq.5: loss function} are the mean squared error (MSE) of the energy per atom, ionic forces, and virial isotropic pressure, respectively, and $W_E$, $W_F$, and $W_P$ are tunable weights multiplying the different MSEs. 
Other functional forms of the loss function are possible, including additional physical quantities (e.g., charges, all $6$ independent components of the stress, etc). In neural networks, $\mathcal{L}(\mathbf{p})$ can be minimized using stochastic optimization methods that compute the gradients using only a small batch of training configurations. 
In this paper, we 
primarily use MLIPs based on the kernel ridge regression (KRR) method. Within KRR, the local atomic energy is expressed as 
\begin{equation}
    e\left( \mathcal{R}; \mathbf{p}\right) = \sum_{\nu = 1}^{N_{\textrm{env}}} p_{\nu} K \left( \mathcal{R}, \mathcal{R}_{\nu}\right) \label{eq.6: KRR}
\end{equation}
where $\mathcal{R}_{\nu}$, $\nu = 1 , \dots , N_{\textrm{env}}$ belong to a subset of all the local environments in the training set, and $K(\mathcal{R}, \mathcal{R_{\nu}})$ is the normalized kernel between $\mathcal{R}$ and $\mathcal{R}_{\nu}$. Following Refs~\citenum{Tirelli2022} and \citenum{Tenti2024}, 
we adopt a kernel based on a variant of the Smooth Overlap of Atomic Positions (SOAP) descriptor~\cite{Bartok2013,De2016}(see App.~\ref{appendix: KRR models}), and we select
the $N_{\textrm{env}}$ local environments using the farthest point sampling method~\cite{Tirelli2022}, 
according to
the distance introduced by the kernel $K$. 
Notice how plugging Eq.\eqref{eq.6: KRR} into the loss function \eqref{eq.5: loss function} and taking the derivatives with respect to the variational coefficients turns 
the minimization problem into
a linear system, which is then solved by the conjugate gradient method. 
We conclude this Section by stressing the fact that, by definition, any MLIPs will give energy derivatives (e.g., forces and virial stress components) that are consistent with their underlying potential energy surface. Any effect of potential inconsistency in the dataset, such as the ones studied in this work, will manifest in the final performance of the models and its dependence
on the relative training weights $W_E$, $W_F$, and $W_P$. This is demonstrated in Sec.\ref{sec: dataset comparison}, where we compare the quality of MLIPs trained on different datasets as a function of $W_E$, $W_F$, and $W_P$. 

\subsection{Self-consistency error in VMC and its correction}\label{sec: SCE correction}

Within VMC, the force acting on the $i$-th atom is computed using
\begin{equation}
\mathbf{F}^{\textrm{VMC}}_i = -\left\langle \frac{\partial}{\partial \mathbf{R}_i}E_L\right\rangle - 2\left\langle (E_L - E) \frac{\partial}{\partial \mathbf{R}_i}\log \Psi \right\rangle , \label{eq: VMC force}
\end{equation}
where $\langle \cdots \rangle$ indicates the quantum expectation value on the wave function $\Psi$ evaluated stochastically, $E_L \equiv \hat{H} \Psi / \Psi$ the local energy with $\hat{H}$ being the Hamiltonian of the system and $E = \langle E_L \rangle$.
To obtain an efficient and statistically meaningful value of the force in Eq.~\eqref{eq: VMC force} with finite variance, techniques such as the zero-variance zero-bias principle~\cite{Assaraf2003}, the space warp transformation~\cite{Umrigar1989} and reweighting~\cite{Assaraf2000,Assaraf2003,Attaccalite2008,Sorella2010,Filippi2016,Nakano2022} are applied. 
$\mathbf{F}^{\textrm{VMC}}_i$ is unbiased only when all the variational parameters of the WF are at their variational minimum. As mentioned earlier, this assumption is not true for partially optimized WFs, such as the frozen-orbital JSD,
one of the most popular choices. 
In this case, an extra term
\begin{equation}
\mathbf{F}^{\textrm{corr}}_i = - \sum_{k = 1}^{N_{SD}} \frac{\partial E}{\partial \alpha^{\textrm{SD}}_{k} }\frac{d \alpha^{\textrm{SD}}_{k}}{d \mathbf{R}_i}\label{eq: VMC force correction}   
\end{equation}
has to be considered, containing the derivatives of the $N_{\textrm{SD}}$ variational coefficients $\alpha_{k}^{\textrm{SD}}$ 
included in the Slater determinant. 

Discarding the term in Eq.\eqref{eq: VMC force correction} causes an inconsistency between the forces and the underlying PES computed with a JSD WF, meaning that $\mathbf{F}^{\textrm{VMC}}_i \neq - \frac{\mathop{dE}}{\mathop{d\mathbf{R}_i}}$. This is called the self-consistency error (SCE)~\cite{Tiihonen2021,Nakano2024}. An analogous discussion applies to other energy derivatives, such as the virial pressure, that contain a similar bias and for which we can derive similar corrections. 

In our applications, following the implementation of the \textsc{TurboRVB} package~\cite{Nakano2020} (here used for all the QMC calculations), we consider a representation of the SD part in terms of the antisymmetrized geminal power (AGP)
\begin{equation}
    \Phi_{\textrm{AGP}} = \mathcal{A} \left( g(x_1 , x_2) \cdots g(x_{N-1}, x_{N})  \right), \label{eq: agp}
\end{equation}
where $x_i = (\mathbf{r}_i , \sigma_i)$ indicates the collective spatial and spin coordinates of the $i$-th electron, with $i = 1, \dots, N$, $\mathcal{A}$ is the antisymmetrizer operator and $g$ is a pairing function. In Eq.~\eqref{eq: agp} we also suppose $N$ to be even for simplicity. In the spin unpolarized case, the pairing function can be expressed as the product of a spatial part $f$ and a spin singlet, namely
\begin{equation}
    g(x_i , x_j ) = f(\mathbf{r}_i, \mathbf{r}_j) \frac{|\uparrow\downarrow\rangle - |\downarrow\uparrow\rangle }{\sqrt{2}}. 
\end{equation}
In \textsc{TurboRVB} the spatial part of the pairing function is written in terms of a localized basis of $L$ atomic orbitals (AOs) $\phi_{k}$ 
\begin{equation}
    f(\mathbf{r}_i, \mathbf{r}_j)  = \sum_{k=1}^L \sum_{k' =1}^L \lambda_{k,k'} \phi_{k}^{\theta} (\mathbf{r}_i) \phi_{k'}^{\theta'} (\mathbf{r}_j). 
\end{equation}
In our case, the AOs are taken as Gaussian type orbitals (GTOs). For periodic systems, such as the ones considered in this work, the localized basis is appropriately generalized to satisfy periodic or twisted boundary conditions~\cite{Dovesi2018}, depending on the phase $\theta$ ($\theta^\prime$) for spin-up (spin-down) electrons. Here, $\theta = -\theta^\prime$, such that the global phase factor is canceled, making the function  $f$ invariant under a global translation.
Fixing this gauge leads to a numerically stable evaluation of the $d \lambda_{k, k'}/d \mathbf{R}$ derivatives implied by Eq.~\ref{eq: VMC force correction}.

An alternative expression for the function $f$ can also be obtained using molecular orbitals (MOs) $\Phi_n (\mathbf{r})= \sum_{k} c_{n,k} \phi_{k}(\mathbf{r} )$. In particular, when the number $M$ of MOs is equal to $N / 2$, the resulting AGP is exactly equivalent to a single SD, \emph{i.e.,} $\Phi_{\textrm{AGP}} \equiv \Phi_{\textrm{SD}}$. 
Within \textsc{TurboRVB}, the MOs (and the corresponding $\lambda_{k, k'}$ matrix) are initialized by a local density approximation~\cite{Perdew1981} (LDA) DFT calculation. This is performed in a basis consisting of the aforementioned AOs with an additional one-body Jastrow factor to automatically take into account Kato cusp conditions~\cite{Kato1957,Nakano2020}. Because of this, the electronic integrals entering into the Kohn-Sham equations are evaluated on a real space grid with a suitably chosen lattice space.
Finally, we also mention that the basis is regularized by removing any eigenvector of the overlap matrix $S_{k,k'} = \langle \phi_k | \phi_{k'} \rangle $ with a corresponding eigenvalue ($s_i$) satisfying the condition $s_i/s_{\rm max} < \epsilon_{\textrm{cut}}$~\cite{Azadi2010}, where $s_{\rm max}$ is the maximum eigenvalue and $\epsilon_{\textrm{cut}}$ is set to $10^{-7}$ in this study. This regularization is very general and is also able to automatically cure the problem of basis-set redundancy coming from delocalized periodic Gaussians.

Finally, the expression in Eq.~\eqref{eq: VMC force correction} 
is stochastically evaluated in VMC according to

\begin{equation}
    \mathbf{F}_i^{\textrm{corr}} = -2 \Re \left\{\left\langle  E_L \sum_{k = 1}^{L} \sum_{k' = 1}^{L} \left[ \left( \mathcal{O}_{k , k'} - \overline{\mathcal{O}}_{k, k'}\right) \frac{d \lambda_{k, k'}}{d \mathbf{R}_i} \right] \right\rangle \right\}\label{eq: force correction stochastic average}, 
\end{equation}
where $\mathcal{O}_{k,k'} = \frac{\partial \log \Psi}{\partial \lambda_{k,k'}}$ and $\overline{\mathcal{O}}_{k, k'} = \langle \mathcal{O}_{k,k'}\rangle$. 
In the previous expression, the logarithmic derivatives in $\mathcal{O}_{k,k'}$ are efficiently computed using the adjoint algorithmic differentiation~\cite{Sorella2010}, while the derivatives of the parameters, $\frac{d \lambda_{k, k'}}{d \mathbf{R}_i}$, 
are evaluated within DFT-LDA 
using the finite differences method (FDM), as described in Ref.~\citenum{Nakano2024}.  
For the virial isotropic pressure, an analogous derivation yields the following expression for the correction term: 

\begin{equation}
   P^{\textrm{corr}} = -2 \Re \left\{\left\langle  E_L \sum_{k = 1}^{L} \sum_{k' = 1}^{L} \left[ \left( \mathcal{O}_{k , k'} - \overline{\mathcal{O}}_{k, k'}\right) \frac{d \lambda_{k, k'}}{d V} \right] \right\rangle \right\}\label{eq: pressure correction stochastic average}. 
\end{equation}

The above Equation can be easily extended to the non-isotropic case.

\section{Results}\label{sec:results}

Henceforth,
we analyze how the SCE affects the training of MLIPs and discuss how the correction described in Sec.~\ref{sec: SCE correction}
can effectively be implemented to improve the accuracy of the resulting models. 
With this aim, we consider a dataset made of pristine hydrogen configurations for which we computed VMC energy, forces, and pressure using different types of WF and with/without bias correction. 
The dataset is the same as the one used in Ref.~\citenum{Tenti2024} and comprises $558$ configurations of $128$ atoms each. These  configurations were extracted from
DFT-driven MD simulations at temperatures between $4000$~K and $20000$~K and Wigner-Seitz radii $r_s$ between $1.80$ and $2.12$. This range of thermodynamic conditions corresponds to the expected location of the Hugoniot curve of deuterium, defined as the set of states that can be reached by producing a shock wave in a sample. For these configurations, we fully optimized the wave function, going beyond the frozen-orbital JSD ansatz.
As previously 
mentioned,
we take this dataset as the reference one.
In Sec.~\ref{sec: bias in Hugoniot dataset}, we study the SCE
for the same set of configurations 
as
the reference dataset,
and show the effect of both forces and pressure corrections. 
In Sec.~\ref{sec: dataset comparison} we directly compare the accuracy of MLIPs trained on a biased dataset (i.e. affected by the SCE) and a corrected one, respectively. We 
also compare both models with the reference
dataset used in Ref.~\citenum{Tenti2024}.
Finally, in Sec.~\ref{sec: MD results}, we 
analyze the results of molecular dynamics simulations driven by the different MLIPs. 

\subsection{Correcting the SCE bias in the Hugoniot dataset}\label{sec: bias in Hugoniot dataset}

To assess the magnitude of the SCE in the Hugoniot dataset, we use a JSD WF with a basis set of [4s2p1d] and [2s2p1d] GTOs for the antisymmetric part and Jastrow factor, respectively. The same primitive basis set was used in Ref.~\citenum{Tenti2024}. We
initialize the SD by running an LDA-DFT calculation with twisted boundary conditions at the Baldereschi point, i.e, at $\mathbf{k} = \left( \frac{1}{4},\frac{1}{4},\frac{1}{4}\right)$ in crystal units. The Jastrow part is optimized using the stochastic reconfiguration method~\cite{Sorella1998}. 

\begin{figure}
     \centering
     \includegraphics[width=1.0\textwidth]{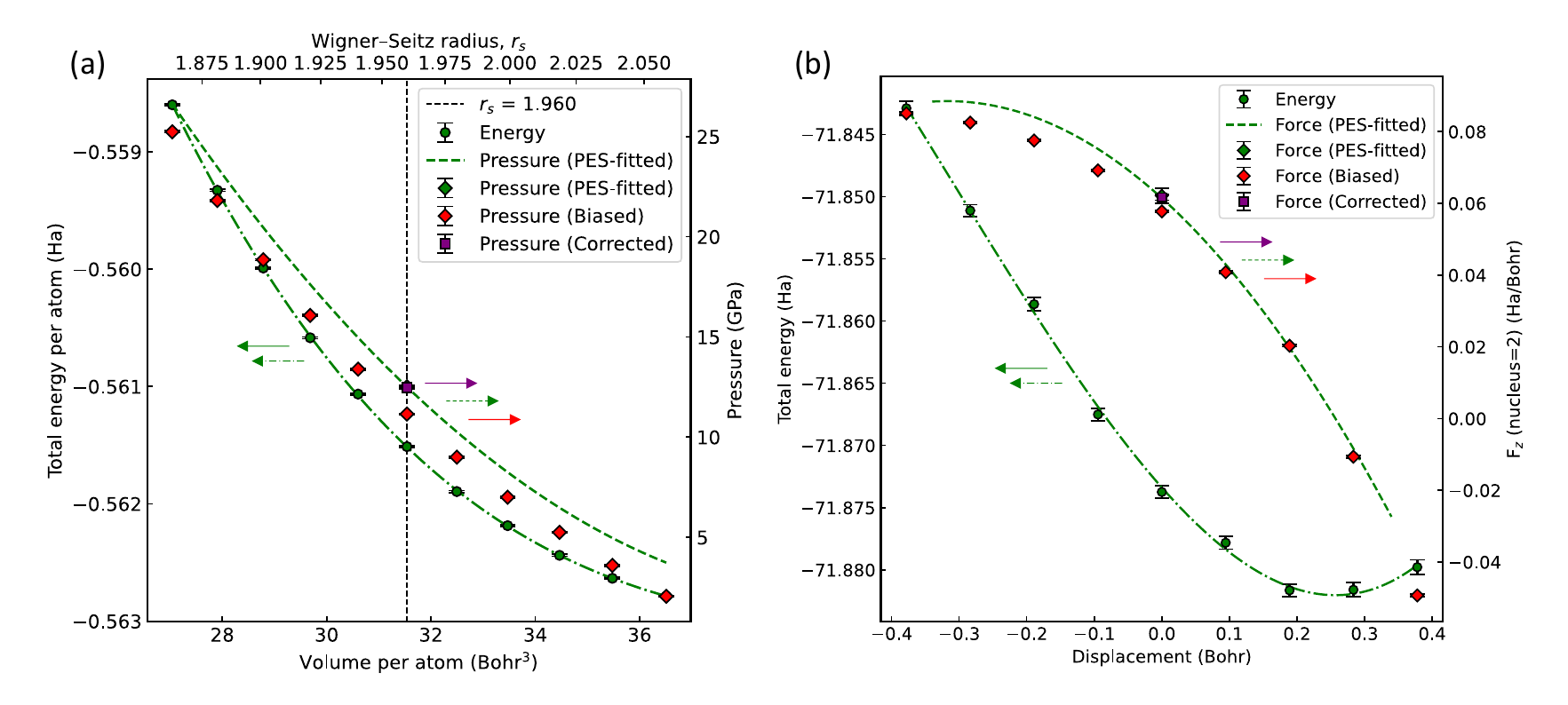}
     \caption{(a) Comparison among the biased pressure evaluated with an expression equivalent to Eq.~\eqref{eq: VMC force} (red diamonds), the corrected pressure obtained by applying Eq.~\eqref{eq: pressure correction stochastic average} (violet square), and the pressure obtained by fitting the PES (dashed green line and green diamond)
     for a selected configuration in the Hugoniot dataset\cite{Tenti2024}. 
     The PES of the system is also shown (green dots and dash-dotted line). (b) Comparison among the biased force evaluated with Eq.~\eqref{eq: VMC force} (red diamonds), the corrected force obtained by applying Eq.~\eqref{eq: force correction stochastic average} (violet square), and the force calculated by fitting the PES (dashed green line and green diamond)
     for the $z$-component of the second hydrogen atom belonging to the same configuration as in (a). 
     The PES of the system along the atomic displacement is also shown (green dots and dash-dotted line).}
     \label{fig: force pressure bias configuration 1}
\end{figure}

We  assess the consistency of the JSD forces and pressures by calculating the forces and pressure on a few selected configurations with two separate approaches: (i) evaluating Eq.\eqref{eq: VMC force} (along with the analogous expression for pressure) and (ii) using the 
FDM by fitting the PES of the system and evaluating the derivative numerically. As previously discussed, the SCE manifests itself as a difference between these two values. 
The results are shown in Fig.~\ref{fig: force pressure bias configuration 1} for one of the force components and the virial pressure. 

\begin{figure}
     \centering
     \includegraphics[width=1.0\textwidth]{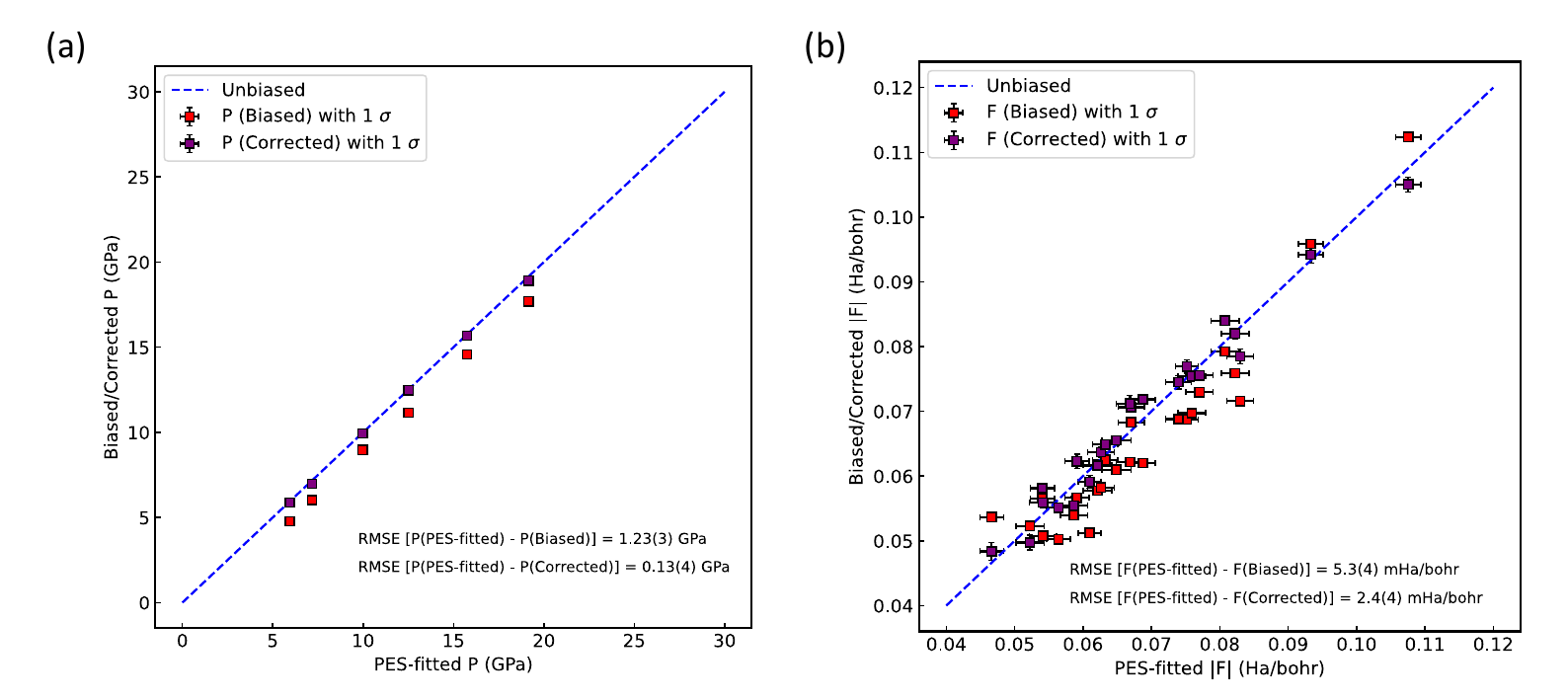}
     \caption{(a) Value of the biased (red markers) and corrected (violet markers) pressure for $6$ different $128$-atom configurations as a function of the numerical pressure estimated from the derivative of the PES with respect to the volume. (b) Values of the biased (red markers) and corrected (violet markers) force components as a function of the numerical force estimated by the PES 
     for one of the configurations belonging to the Hugoniot dataset\cite{Tenti2024}.
     The dashed line 
     is a reference for
     perfect consistency.}
     \label{fig: force pressure bias summary}
\end{figure}

We 
observe a significant bias for both quantities, with a discrepancy as large as $\sim 10\%$ of their values. This clearly highlights the importance of curing the SCE for the 
biased dataset. 
For this reason, we applied the corrections of Eqs.~\eqref{eq: force correction stochastic average} and \eqref{eq: pressure correction stochastic average} to forces and pressures, respectively. For forces, we computed the numerical derivatives of the parameters $\frac{d \lambda_{k, k'}}{ d \mathbf{R}}$ with the FDM from DFT-LDA, the theory used to obtain $\lambda_{k, k'}$, 
with a displacement $\Delta R = \pm 0.003$~Å. The derivatives of $\lambda_{k, k'}$ with respect to the volume, entering in the pressure correction, were obtained with the FDM using relative volume variations of $\pm 0.03 \%$. 
In Fig.~\ref{fig: force pressure bias configuration 1} we 
observe that the corrected forces and pressure are perfectly compatible with their values 
directly computed through the fit of the VMC PES. 
Fig.~\ref{fig: force pressure bias summary} shows the values of both the biased and corrected force 
components (pressures) versus the forces (pressures) estimated from the PES, for several configurations.
Notice how the pressure correction acts almost like a rigid shift of $\sim 1$~GPa.
The root mean squared error (RMSE) of
biased and corrected quantities is also shown, further demonstrating the effectiveness of the correction in removing the SCE. 

Sometimes, the added correction significantly increases the error bar $\sigma_F$ of the corrected forces. We then
disregard the corrections that lead to forces with $3 \sigma_F > 0.015$ Ha/Bohr. For pressures, the threshold value of $1.5 \times 10^{-5}$ is
applied to $3 \sigma_p$. See App.~\ref{appendix: cutoff} for further details. 
In principle, the FDM used to estimate the parameter derivatives $\frac{d \lambda_{k, k'}}{ d \mathbf{R}}$ can be replaced with more robust and faster approaches (\emph{e.g.}, using linear response theory~\cite{Toulouse2018}). We leave this possibility for future work.

\subsection{Comparison between VMC datasets}\label{sec: dataset comparison}

To illustrate how 
using an SCE-corrected JSD dataset affects the MLIPs generation, we compare the performances of MLIPs trained on three different datasets: (i) the aforementioned "biased dataset" obtained with a JSD WF and with forces and pressure computed with the standard expressions (Eq.~\ref{eq: VMC force}), (ii) a "corrected dataset" where the corrections of Eqs.~\eqref{eq: force correction stochastic average} and \eqref{eq: pressure correction stochastic average} 
are applied as described in Sec.~\ref{sec: SCE correction}
and (iii) a "reference dataset", \emph{i.e.}, the one introduced in Ref.~\citenum{Tenti2024}. In particular, the latter was obtained from an improved Jastrow-correlated AGP WF, where the Jastrow and antisymmetric part were both optimized, thanks to a combination of geminal embedded orbitals\cite{Sorella2015} to reduce the number of variational parameters and a restricted optimization based on the locality of the AGP $\lambda_{k, k'}$ matrix. 
The $\lambda_{k, k'}$ coefficients 
were optimized for those basis set elements $k$ and $k'$ centered on atoms closer than a cutoff distance of 4 Bohr radii\cite{Tenti2024}.
No significant lowering of the variational energy 
was found beyond this threshold, suggesting that the dependence of the energy on the remaining parameters can be disregarded. 
Thanks to this approach,
the SCE on this dataset was effectively removed and we thus decided to take the models trained on it as the reference. 
Notice how the explicit optimization of the whole WF, as the one performed in Ref.~\cite{Tenti2024}, is a procedure that is about an order of magnitude more expensive than the partial optimization of the frozen-orbital JSD WF,
and requires advanced minimization techniques~\cite{Holzmann2003,Slootman2024}. 

For all the MLIPs, a KRR model
is trained on the difference between VMC  quantities and DFT ones, the latter using the Perdew-Burke-Ernzerhof (PBE) exchange-correlation functional~\cite{Perdew1996}, following the $\Delta$-learning approach~\cite{Ramakrishnan2015,Tirelli2022,Tenti2024}. 
In this way, a much higher accuracy 
is achieved with relatively small datasets such as the ones used here. The values of the hyperparameters chosen for these models are reported in App.~\ref{appendix: KRR models}. 
Among the $558$ configurations of the datasets, we select $58$ configurations for testing.
We investigate the effect of varying the training weights in the loss function (Eq.~\eqref{eq.5: loss function}) on the RMSE 
computed for energy, forces, and pressure of the test set for each dataset.

Since our goal here is to compare MLIPs trained on different datasets (\emph{i.e.}, different 
training
values for energy $E$, forces $\mathbf{F}$ and pressure $P$),  
the RMSE may not be the best metric to assess the relative accuracy of
one model with respect to the others. For this reason, we define the relative improvement of the model for each system property ($X = E, \mathbf{F}, P$) as: 

\begin{equation}
    \Delta_X = \frac{ \mathrm{RMSE}(X_{\textrm{test}} - X_{\textrm{dummy}}) - \mathrm{RMSE}(X_{\textrm{test}} - X_{\textrm{pred}}) }{\mathrm{RMSE}(X_{\textrm{test}} - X_{\textrm{dummy}})}, \label{eq: Delta models}
\end{equation}
where $X_{\textrm{pred}}$ are the model predictions, $X_{\textrm{test}}$ the test set values, $E_{\textrm{dummy}} = \frac{1}{N_{test}}\sum_{i \in test} E_i$, $\mathbf{F}_{\textrm{dummy}} = 0$, and  $P_{\textrm{dummy}} = 0$.  
In other words, $\Delta_X$ measures the relative improvement of the model prediction on the quantity $X$ with respect to a "dummy" model, corresponding to a perfectly flat PES equal to the average energy on the test set for all configurations. Notice how $\Delta_X = 1$ when $\mathrm{RMSE}(X_{\textrm{test}} - X_{\textrm{pred}}) = 0$, indicating that the model has learned exactly the quantity $X$, while a value $\Delta_X \leq 0$ means that the model has an error equal to, or larger than, a flat model. 
In the present work, 
the test quantities refer to the difference between the VMC and the DFT-PBE baseline, while $E_{\textrm{dummy}}$ is the average difference on the test set.

\begin{figure}
    \centering
    \includegraphics[width=\linewidth]{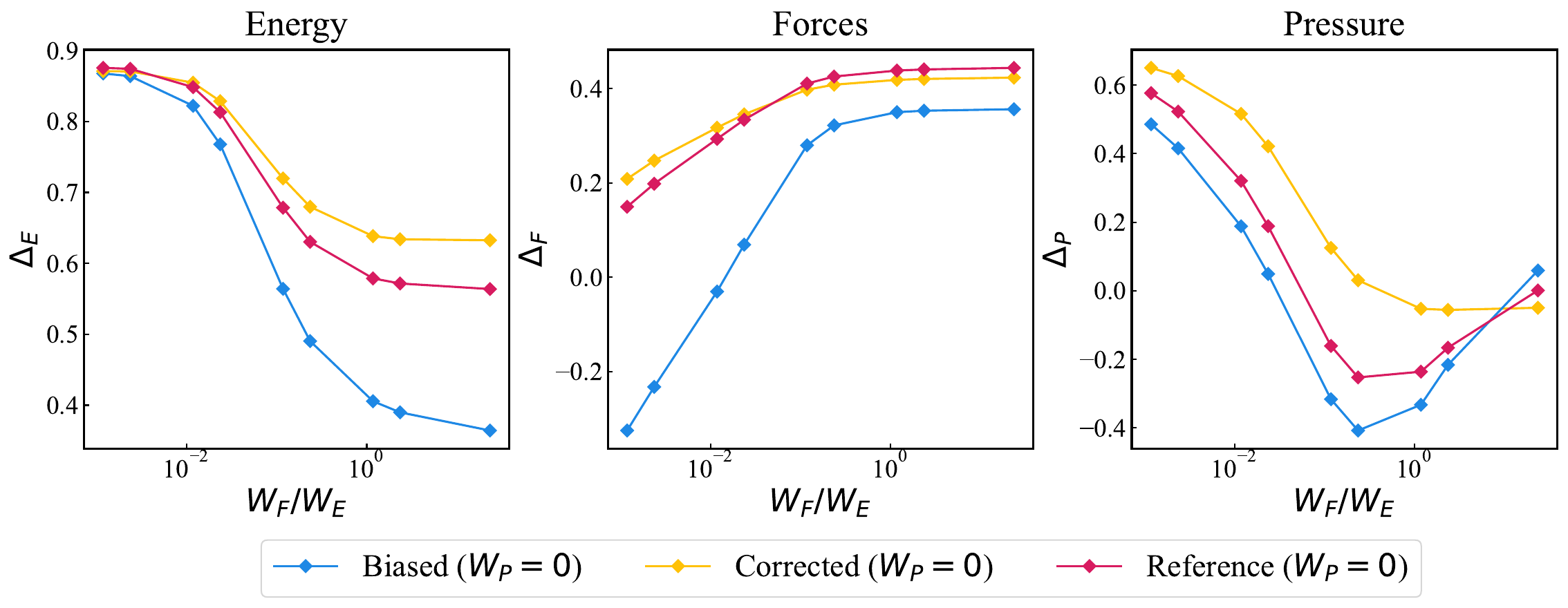}
    \caption{Relative RMSE variation $\Delta_{X}$ 
    (Eq.~\ref{eq: Delta models}) for energy, forces, and pressure, as a function of the force/energy weight ratio in the loss function of Eq.~\ref{eq.5: loss function} (expressed in 
    atomic units) for MLIPs trained on the biased, corrected and reference datasets, respectively. The weight on the pressure 
    is set to zero for all models.}
    \label{fig: relative rmse models opt-unopt-corr no pressure}
\end{figure}

\begin{figure}
    \centering
    \includegraphics[width=\linewidth]{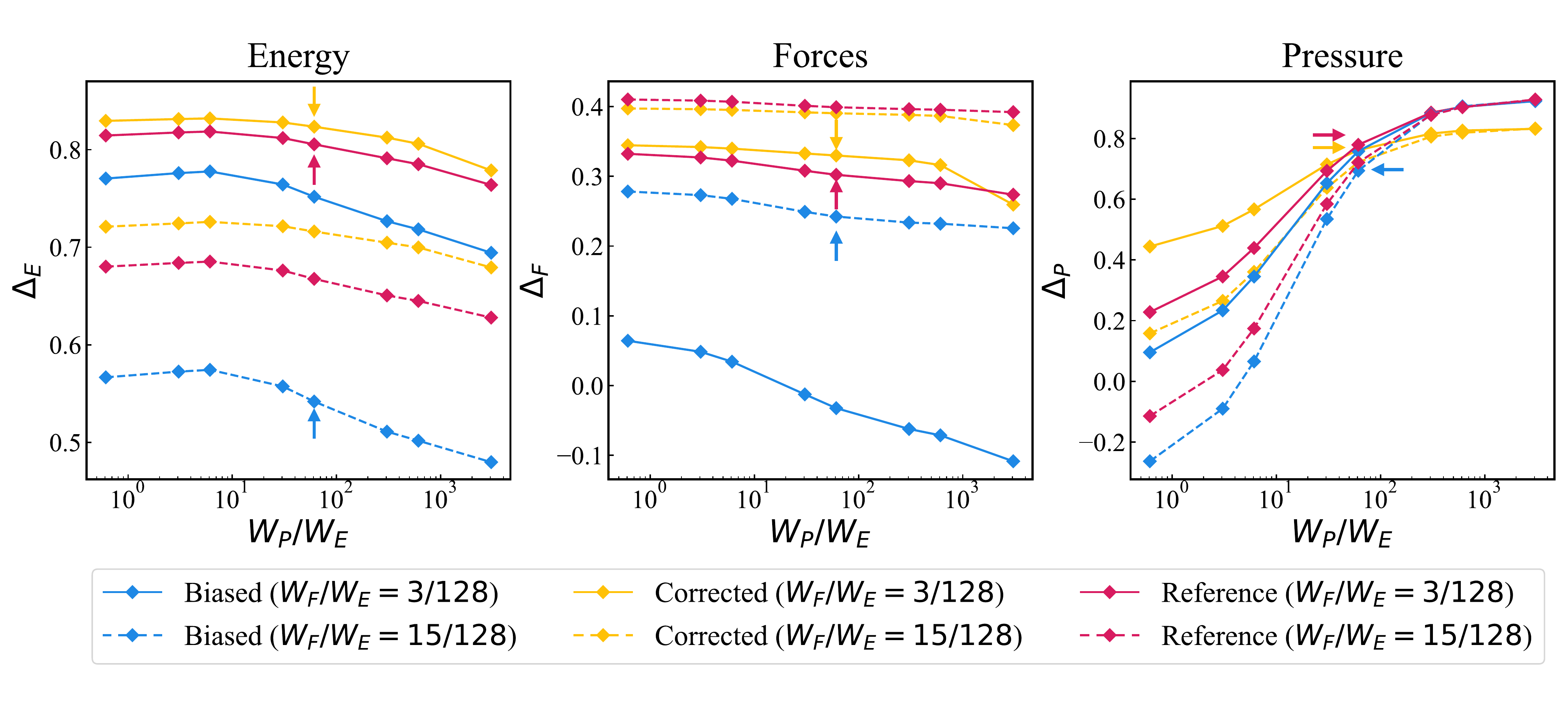}
    \caption{$\Delta_{X}$ 
    (Eq.~\ref{eq: Delta models})
    for energy, forces, and pressure, as a function of the pressure/energy weight ratio in the loss function of Eq.~\ref{eq.5: loss function} used for training (expressed in 
    atomic units), for the different models. Solid lines correspond to $W_E = 1$ and $W_F = 3 / 128$, while dashed lines correspond to $W_E = 1$ and $W_F = 15 / 128$. The arrows highlight the "best" MLIPs for each of the three datasets, which are
    used for the MD simulations described in Sec.~\ref{sec: MD results}.}
    \label{fig: relative rmse models opt-unopt-corr pressure}
\end{figure}

Figs.~\ref{fig: relative rmse models opt-unopt-corr no pressure} and \ref{fig: relative rmse models opt-unopt-corr pressure} show the results obtained for $\Delta_E$, $\Delta_F$, and $\Delta_P$ using MLIPs trained with different values of the training weights $W_E$, $W_F$, and $W_P$.
We first analyze what happens by progressively increasing the weight on the forces for the different models without explicitly training on pressures (\emph{i.e.}, $W_P = 0$). The force bias in the biased dataset is evident from the lower $\Delta_F$ of the corresponding model compared to that of both the corrected and reference datasets (see Fig.~\ref{fig: relative rmse models opt-unopt-corr no pressure}). Notice how, for small values of $W_F / W_E$, $\Delta_F$ is negative in the biased case, which in the $\Delta$-learning framework means that the correction on forces is detrimental. 
As the weight on forces is increased, all models reach a plateau in $\Delta_F$, while simultaneously losing some accuracy on the other quantities. For the energy, this loss of accuracy is significantly larger in the models trained on the biased dataset, further demonstrating the effect of the SCE.

We observe very similar performances between the corrected and reference models. For pressure, the models including the SCE correction
show the best performances for almost all the values of $W_F$. Since the loss does not include the pressure in this case, this is another indication of consistency. 
In Fig.~\ref{fig: relative rmse models opt-unopt-corr pressure} 
we plot the results obtained by varying the weight on pressure, 
for two values of $W_F/W_E$, namely $W_F/W_E = 3/ 128$ and $W_F/W_E = 15/ 128$. Also in this case, a better accuracy is generally achieved for the models trained on the corrected and reference datasets.

Finally, we mention how these conclusions exclusively depend on the quality of the datasets themself, \emph{i.e.}, their internal consistency, and not on the MLIPs specific architecture. To demonstrate this feature, in App.~\ref{appendix: MACE dataset comparison}, we discuss 
the relative improvement $\Delta_X$ for models obtained with the MACE~\cite{Batatia2022} framework, which show the same behavior as the one found for the KRR MLIPs.

\subsection{MD results comparison}\label{sec: MD results}
To conclude our analysis, we 
run a set of MD simulations using MLIPs trained on the three different datasets. For each model, we select the relative training weights $W_F / W_E$ and $W_P / W_E$ to have a good tradeoff between the RMSE values on energy, forces, and pressure. In particular, for the MLIPs trained on the reference and corrected weight, we choose the last value of $W_F / W_E$ before the significant decrease in $\Delta_E$ (\emph{i.e.}, the fourth smallest value in Fig.~\ref{fig: relative rmse models opt-unopt-corr no pressure}), in order to increase 
as much as possible 
the accuracy on forces without spoiling too much the energy. A similar reasoning is used to determine the optimal value of $W_P / W_E$ (see Fig.~\ref{fig: relative rmse models opt-unopt-corr pressure}). 
For the MLIP trained on the biased dataset, we 
choose a larger value of $W_F / W_E$, which 
is necessary to reach a reasonable accuracy on forces. This comes at the price of sacrificing the accuracy on the energy, a direct consequence of the SCE. 
The weights used for each MLIP are summarized in Tab.~\ref{tab: training weights}. While the values of the best relative training weights $W_F / W_E$ and $W_P / W_E$ are model dependent, we highlight that this procedure is generally applicable to any other system.

\begin{table}[]
    \centering
    \begin{tabular}{l|cc}
    \toprule
    Dataset & $W_F / W_E$ & $W_P / W_E$ \\
    \midrule
        Biased & $15 / 128$ & $10^6 / 128^2$ \\
        Corrected & $3 / 128$ & $10^6 / 128^2$ \\
        Reference & $3 / 128$ & $10^6 / 128^2$ \\
    \bottomrule
    \end{tabular}
    \caption{Training weights for the MLIPs used in the MD simulations, for each dataset. For both the forces and pressure weights, we report the relative value with respect to $W_E$. 
    All values are in atomic units.}
    \label{tab: training weights}
\end{table}

We run MD simulations at three different 
densities and temperature conditions
close to the position of the deuterium Hugoniot curve given by the VMC-MLIP of Ref.~\citenum{Tenti2024}. In particular, we considered thermodynamic conditions that span both the molecular liquid and the atomic one.
Because of the $\Delta$-learning approach adopted, the resulting efficiency of these calculations is the same as a DFT Born-Oppeheimer MD simulation. Indeed, the energy, ionic forces, and pressure of the system are calculated at each step with DFT, and the corresponding corrections are then added using the MLIP. Notice how the DFT baseline can be, in principle, replaced by a faster potential (\emph{e.g.}, using an MLIP trained on DFT data~\cite{Tenti2025}), improving the overall efficiency.
Further details regarding the computational 
setup
of the MD 
simulations
are given in App.~\ref{appendix: MD details}. 

\begin{figure}
    \centering
    \includegraphics[width=0.85\linewidth]{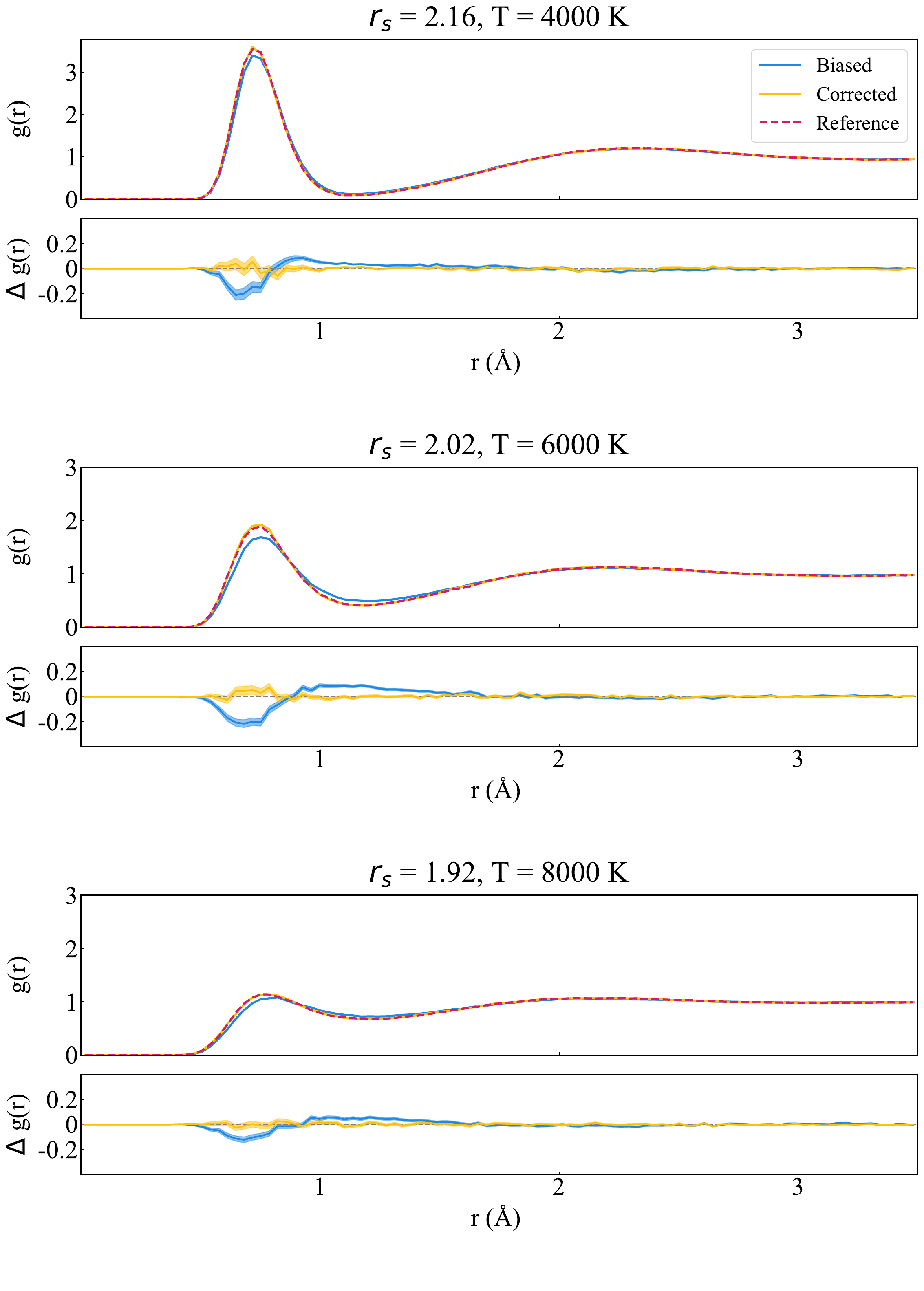}
    \caption{Radial distribution function $g(r)$ for densities and temperatures close to the Hugoniot curve,
    obtained with MLIPs trained on different datasets. The difference $\Delta g(r)$ with respect to the model trained on the reference dataset is also shown for both the "biased" and "corrected" models. The shaded area indicates the 
    statistical
    uncertainty in $\Delta g(r)$. }
    \label{fig: g(r) comparison models}
\end{figure}

A comparison of the equilibrium pressure during the dynamics for three different temperatures is reported in Tab.~\ref{tab: pressure comparison models (biased , optimized , corrected)}. 
Notice that, in principle, the reference dataset and the corrected one can give different results. Indeed, the optimization of the antisymmetric part of the WF not only improves the consistency of forces and pressure, but can also modify the PES of the system. 
Nevertheless, if compared with the biased model, the MLIP trained on the corrected data gives results that are 
consistently much closer to the reference ones. This suggests that
the main discrepancy between the biased and reference models comes from the SCE, and that the SCE correction improves considerably the physical description of the system. 
We observe a maximum discrepancy of $\sim 2$~GPa at $r_s = 2.02$ and $T=6000$~K
between the biased and corrected results, while the corrected and reference ones are in statistical accordance within a joint error bar of 0.2 GPa at the same thermodynamic conditions.   
Remarkably, this state is very close to the Hugoniot compressibility peak, \emph{i.e.}, the maximum in the density reached by the shocked state along the Hugoniot curve. This 
thermodynamic point
is very sensitive to changes in the underlying equation of state, and a pressure shift as the one observed here between the corrected and biased models has a significant impact in the final position of the Hugoniot~\cite{Clay2019}. 
Similar conclusions can be reached by looking at the radial distribution function $g(r)$ (Fig.~\ref{fig: g(r) comparison models}). 
Analogously to the pressure analysis,
the $g(r)$ of the corrected model is compatible with the one given by the reference one.
Also for this quantity the maximum discrepancy between the corrected and biased models is observed close to the compressibility peak at $T = 6000$~K, which corresponds to the onset of the the molecular-to-atomic transition along the Hugoniot curve. 
These results further highlight how SCE forces and pressure 
corrections significantly improve the physical outcome of the resulting MLIPs. 

\begin{table}[]
    \centering
    \begin{tabular}{l|ccc}
    \toprule
         & $P_{\textrm{bias}}$ (GPa)& $P_{\textrm{corr}}$ (GPa)& $P_{\textrm{ref}}$ (GPa)\\
         \midrule
        $r_s = 2.16$, $T=4000$~K  & $17.1(1)$ &  $17.4(1)$ & $17.4(1)$\\
        $r_s = 2.02$, $T=6000$~K & $26.8(1)$& $28.5(1)$ & $28.7(1)$\\
        $r_s = 1.92$, $T=8000$~K   & $38.3(1)$ & $39.8(1)$& $39.9(1)$\\
         \bottomrule
    \end{tabular}
    \caption{Average pressure from MD simulations obtained with different MLIPs trained on the biased, corrected,
    and reference datasets, respectively. All models employ the $\Delta$-learning technique with a DFT-PBE baseline. }
    \label{tab: pressure comparison models (biased , optimized , corrected)}
\end{table}

\section{Conclusions}

Machine learning models have 
become a key tool to reach system sizes and simulation times not achievable with first principle methods, while showing a similar accuracy.
In particular, MLIPs 
trained on data generated by
advanced quantum-mechanical algorithms, such as quantum Monte Carlo (QMC), are essential to improve the description and our physical understanding of phenomena for which standard mean-field methods fail.
In this work, we 
provided evidence on
how VMC, in combination with a cheap-to-optimize JSD WF and a force/pressure correction, can be used to construct 
reliable
QMC-based training sets. 
Our results can be summarized in the following three points.

First, we clearly demonstrated how the SCE affecting derivative quantities (\emph{i.e.}, forces and pressures) computed 
with a partially optimized WF, such as the
frozen-orbital JSD, 
can heavily spoil the accuracy of MLIPs trained on those data. This is manifested by a large dependence of the models' accuracy on the relative weights 
in the loss function. 

Second, we showed how 
an easy-to-evaluate bias correction, previously developed in Ref.~\citenum{Nakano2024}, can be applied to remove the SCE 
from the training set,
and improve the accuracy of the resulting MLIPs.

Finally, 
we directly compared equilibrium thermodynamic quantities 
yielded by MD simulations driven by different MLIPs 
for the deuterium Hugoniot curve, taken as test system.
We observed how the main discrepancy between MLIPs obtained from frozen-orbital and fully optimized WF datasets does not necessarily come from a different PES description of the dataset but from the intrinsic SCE affecting forces and pressures computed with partially optimized WFs.
Indeed, in our working case 
provided by high-pressure hydrogen, the MLIPs trained on the corrected dataset give results in statistical agreement with those 
yielded by
a reference model trained on VMC data 
with
a fully optimized WF. 

Our results 
reveal
the importance of appropriately curing forces and pressures when they are computed with partially optimized WF in QMC and subsequently used to train MLIPs. The MLIPs quality degradation due to the SCE plaguing these VMC datasets is an effect that has been overlooked in previous applications. 
This could be even more relevant for heavier elements, since the SCE usually grows with the number of valence electrons~\cite{Tiihonen2021}. Moreover, the SCE correction scheme is general and applicable to systems more complex than hydrogen, because it is based on cheaper DFT calculations. We have also shown that some numerical instabilities coming from the finite difference approximation used to evaluate the SCE correction can be cured by an appropriate cutoff in the corresponding error bars. The development of a more efficient scheme for evaluating the SCE correction based on linear response theory is currently in progress.
The availability of an SCE correction framework used in combination with other approaches, such as $\Delta$-learning, to further increase the precision of the final models, opens the way to the efficient generation of accurate MLIP based on QMC
electronic structure calculations.

\section*{Acknowledgments}
K.N. is grateful for computational resources from the Numerical Materials Simulator at National Institute for Materials Science (NIMS), and from Research Institute for Information Technology at Kyushu University under the category of General Projects on the supercomputer GENKAI.
M.C. acknowledges GENCI for providing computational resources on the CEA-TGCC Irene supercomputer cluster under project number A0170906493 and the TGCC special session.
M. C. is grateful to EuroHPC for the computational grant EHPC-EXT-2024E01-064 allocated on Leonardo booster partition.

K.N. acknowledges financial support from MEXT Leading Initiative for Excellent Young Researchers (Grant No.~JPMXS0320220025), from Murata Science and Education Foundation (Grant No.~M24AN006), and from Japan Science and Technology Agency (JST), PRESTO (Grant No.~JPMJPR24J9).
M. C. thanks the European High Performance Computing Joint Undertaking (JU) for the partial support through the "EU-Japan Alliance in HPC" HANAMI project 
(Hpc AlliaNce for Applications and supercoMputing Innovation: the Europe - Japan collaboration). 

\section*{Associated Content}

\subsection*{Data Availability Statement}
The {\textit{ab initio}} QMC package used in this work, TurboRVB, is available from its GitHub repository \url{https://github.com/sissaschool/turborvb}.

All data necessary for reproducing the results reported in this work is available in the following GitHub repository \url{https://github.com/giacomotenti/unbiased_qmc_ML}

\appendix
\section{Appendix}

\subsection{Kernel regression models parameters}\label{appendix: KRR models}

In this Appendix we provide further details on the hyperparameters and specific kernels used for the models described in Sec.~\ref{sec: dataset comparison}. The KRR models use a kernel as the one described in Ref.~\citenum{Tirelli2022}. Each environment is described by a density function written as a sum of Gaussian functions (with spread $\epsilon$) multiplied by a localized function $f_c$ depending on a cutoff radius $r_c$, \emph{i.e.,}
\begin{equation}
    \rho\left(\mathbf{r}, \mathcal{R}_i \right) \propto \sum_{ \left|\mathbf{R}_{ij} \right| \leq r_c} f_c \left(\left|\mathbf{R}_{ij}\right| \right) \exp\left( - \frac{\left|\mathbf{r} - \mathbf{R}_{ij}\right|^2}{\epsilon}\right) \label{eq: density}. 
\end{equation}

A similarity kernel between $\mathcal{R}_i$ and $\mathcal{R}_j$ is then defined by averaging the overlap between the two 
densities
over a discrete set of $N_{\textrm{sym}}$ symmetry operations $\hat{U}_k$: 

\begin{equation}
    \mathcal{K}\left(\mathcal{R}_i,\mathcal{R}_j \right) = \frac{1}{N_{\textrm{sym}}}\sum_{k = 1}^{N_{\mathrm{sym}}} \left[\int d^3 \mathbf{r} \rho(\mathbf{r}; \mathcal{R}_i) \rho(\mathbf{r}; \hat{U}_k\mathcal{R}_j) \right]^n. \label{eq: similarity kernel}
\end{equation}

Finally, the normalized kernel is written as
\begin{equation}
        K \left( \mathcal{R}_i , \mathcal{R}_j\right) = \left[\frac{\mathcal{K}\left( \mathcal{R}_i , \mathcal{R}_j\right) }{\mathcal{K}\left( \mathcal{R}_i , \mathcal{R}_i\right) \mathcal{K}\left( \mathcal{R}_j , \mathcal{R}_j\right)}\right]^{\eta}\label{eq: normalized kernel}. 
\end{equation}

In the application presented here, we chose $r_c = 4$ Bohr and $\epsilon = 1.5$ Bohr$^2$ in Eq.~\eqref{eq: density}. In Eq.~\eqref{eq: similarity kernel} we averaged over the cubic symmetry group (here appropriate since we are using cubic cells) and took $n =2$. In Eq.~\eqref{eq: normalized kernel} a value of $\eta = 2$ was employed. 

For the local energy (Eq.~\eqref{eq.6: KRR}) we also added an additional pair-wise term as done in Ref.~\cite{Tenti2024}, using cubic spline functions~\cite{Xie2023} defined on $10$ equally spaced grid points between $0.3$ Bohr and $5.5$ Bohr.  

Finally, the furthest point sampling method was used to select $N_{\textrm{env}} = 6000$ local environments among the training configurations. 

\subsection{Cutoff in the correction error}
\label{appendix: cutoff}

\begin{figure}
    \centering
    \includegraphics[width=0.8\linewidth]{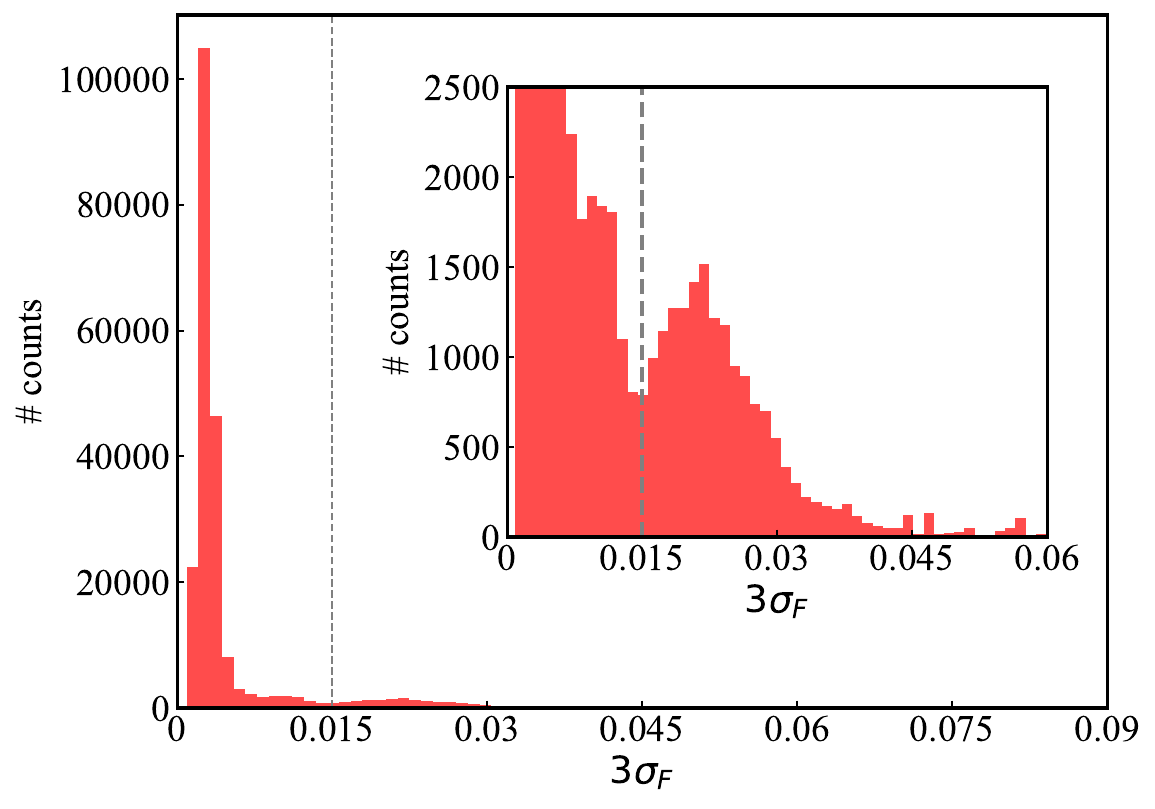}
    \caption{Distribution of three standard deviations for the corrected forces. 
    The inset zooms-in its tail, to highlight
    the outliers. The vertical line indicates the $3\sigma$ threshold chosen in our application, \emph{i.e.}, $0.015$ Ha/Bohr.}
    \label{fig: 3sigma distribution bias correction}
\end{figure}

Figure~\ref{fig: 3sigma distribution bias correction} shows the error bars of the corrected forces in the dataset. We observed that, for certain force components and pressures ($\sim 10\%$ of all components), the correction is occasionally accompanied by large error bars. We identified the cause of this phenomenon in our current workflow, which estimates the parameter derivatives $\cfrac{d \lambda_{k, k'}}{d \mathbf{R}i}$ and $\cfrac{d \lambda_{k, k'}}{d V}$ in Eq.\ref{eq: force correction stochastic average} and Eq.\ref{eq: pressure correction stochastic average} using the FDM by performing multiple DFT calculations at displaced coordinates~\cite{Nakano2024}.
On the one hand, when employing localized basis sets in DFT, a cutoff is usually applied to reduce basis redundancy by eliminating elements corresponding to overlap matrix eigenvalues below a certain threshold, as written in Sec.~{\ref{sec: SCE correction}}.
Without this cutoff, the forces and pressures computed according to Eq.~\ref{eq: VMC force} exhibit large error bars, as reported in Ref.\citenum{Nakano2021}. On the other hand, applying this cutoff causes the basis set to depend on atomic displacements in the FDM approach, as the quality of the basis set changes when ions move. This displacement-induced dependence occasionally deteriorates the error cancellation in Eqs.~\ref{eq: force correction stochastic average} and \ref{eq: pressure correction stochastic average}, resulting in larger error bars.
Another source of large error bars is the finite smearing parameter used for molecular orbital (MO) occupations. Since hydrogen exhibits metallic behavior depending on its density, a smearing technique is necessary to achieve stable and smooth convergence for all configurations. However, employing smearing methods can introduce discontinuities when evaluating parameter derivatives $\cfrac{d \lambda_{k, k'}}{d \mathbf{R}i}$ and $\cfrac{d \lambda_{k, k'}}{d V}$ by FDM.
Additionally, the discrete real-space grid size used in DFT calculations contributes to error deterioration. Within the FDM approach, we need to perform 3$N$ DFT calculations to compute the correction terms. 
To keep their computational cost low
for large $N$, we chose a real-space grid size of $0.15^3$ Bohr$^3$. We underline that the finite-difference step in ionic positions and cell volume for FDM, the smearing parameter, and the real-space grid have been chosen to provide converged forces and pressures, such that any possible bias is comparable with their total stochastic error bars. Yet, the occurrence of large error bars in the SCE correction could not be avoided for some particular ionic configurations by playing solely with these parameters.
These factors indicate that the FDM method proposed in Ref.~\citenum{Nakano2024} can be improved for metallic systems in the future, although it is already 
robust for insulating systems with band gaps, as we have verified for c-BN, SiC, Si, and diamond.

\begin{figure}
    \centering
    \includegraphics[width=\linewidth]{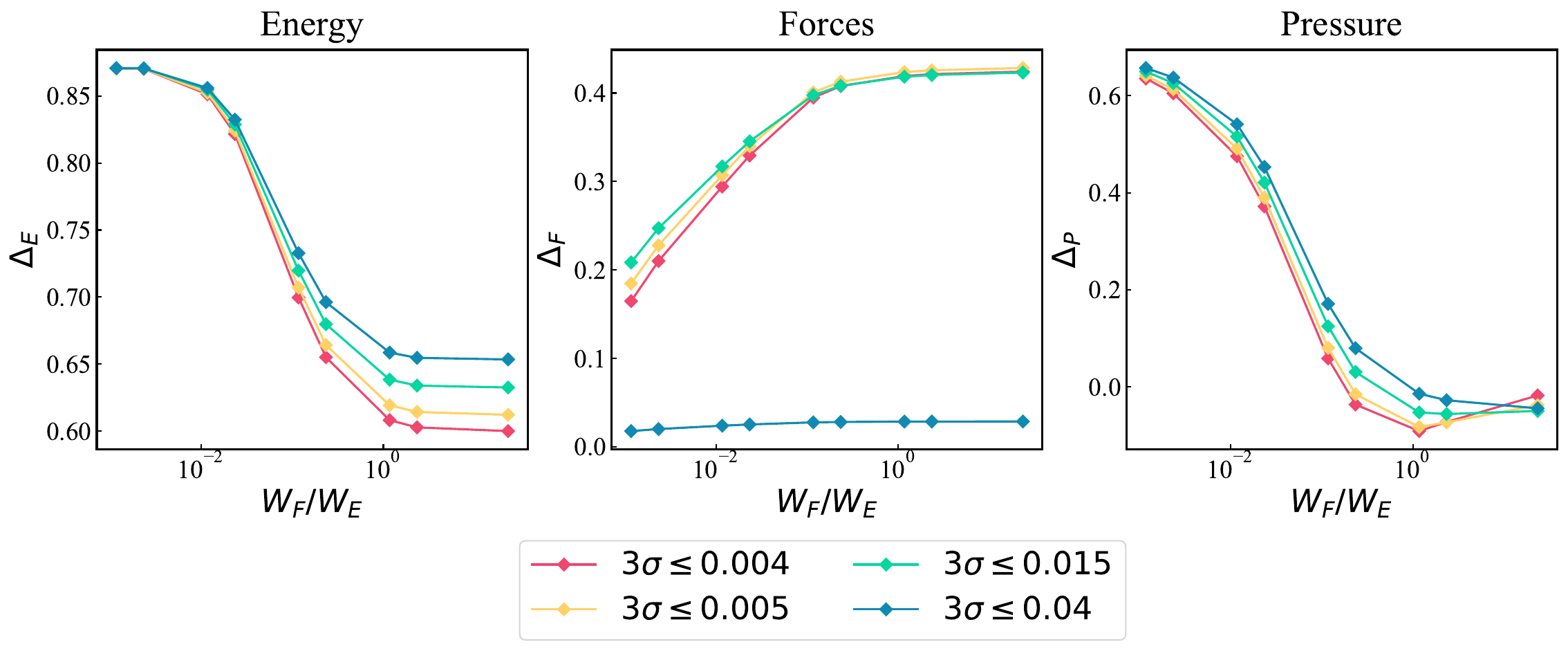}
    \caption{$\Delta_{X}$ for energy, forces, and pressure, as a function of the
    force/energy
    weight ratio in the loss function used for training (expressed in
    atomic units), for models trained on different versions of the corrected dataset. Each dataset corresponds to a different value of the $3 \sigma_F$ threshold used to accept/discard the correction. The thresholds in the legend are reported in Ha/Bohr. For pressure, a threshold of $1.5 \times 10^{-5}$ a.u. 
    is used. The weight on the pressure was set to zero for all models. }
    \label{fig: comparison no pressure vs threshold}
\end{figure}

To manage these erratic components, in this study, we retained the biased force values in the corrected data set whenever three standard deviations exceeded a given threshold, set here to $0.015$~Ha/Bohr. As shown in Fig.~\ref{fig: 3sigma distribution bias correction}, this threshold corresponds to a minimum in the standard deviation distribution for the corrected forces, indicating the onset of departure from normality. A similar strategy was employed for pressure, using a threshold of $1.5 \times 10^{-5}$ a.u.
We confirmed that forces and pressures with error bars below the threshold are unbiased, consistent with their corresponding potential energy surfaces (PESs), as illustrated in Figs.\ref{fig: force pressure bias configuration 1} and \ref{fig: force pressure bias summary}. This choice was further validated by analyzing the behavior of $\Delta_E$, $\Delta_F$, and $\Delta_P$ (see Eq.\eqref{eq: Delta models}) as a function of the relative weight $W_F / W_E$ in the loss function, using models trained on datasets corrected with different thresholds for $\sigma_F$. The results, depicted in Fig.~\ref{fig: comparison no pressure vs threshold}, confirm that the threshold of $0.015$ Ha/Bohr yields the best overall performance. Indeed, although $\Delta_E$ and $\Delta_P$ both steadily increase with increasing threshold values (for fixed $W_F / W_E$), the accuracy of forces deteriorates significantly when corrections with large error bars (corresponding to 
the largest threshold, i.e.
$3\sigma_F \leq 0.04$ Ha/Bohr) are included.

\subsection{MACE models comparison}\label{appendix: MACE dataset comparison}

\begin{figure}
    \centering
    \includegraphics[width=\linewidth]{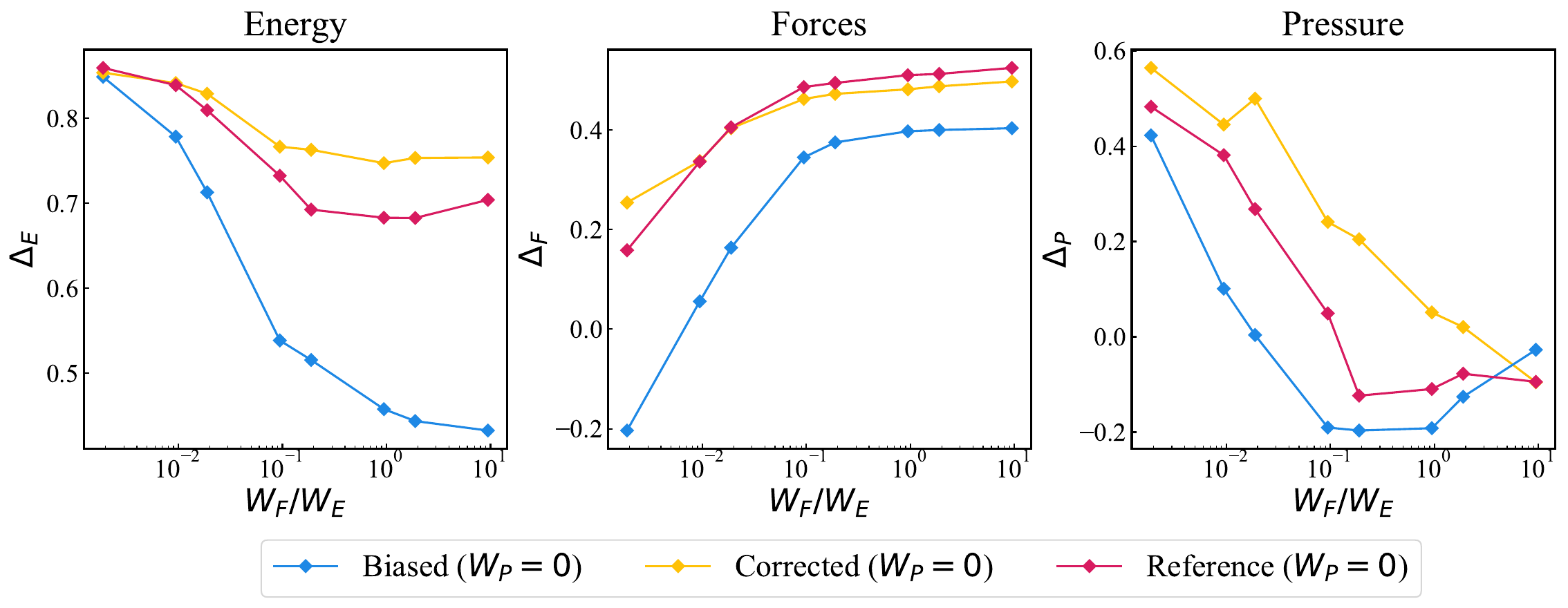}
    \caption{Relative RMSE variation $\Delta_{X}$ 
    (Eq.~\ref{eq: Delta models})
    for energy, forces, and pressure, as a function of the force/energy weight ratio in the loss function (expressed in 
    atomic units) for MACE MLIPs trained on the biased, corrected and reference datasets, respectively. The weight on the pressure 
    is set to zero for all models.}
    \label{fig: comparison no pressure MACE}
\end{figure}

Here, we further support the conclusions made in Sec.~\ref{sec: dataset comparison} by carrying out the same analysis using 
MACE~\cite{Batatia2022}. This is
a completely different MLIP architecture, because it is a framework based on equivariant message-passing neural networks with high-body order messages.
The scope of this 
complementary analysis
is to demonstrate that the behavior observed with the KRR models is general and only due to the training set consistency, and not linked to any particular implementation of the MLIP. 
The models
consider
$256$ inner invariant features and a cutoff of $3$~Å (effectively doubled when considering the message passing step). For each dataset, both training and test sets are the same as the one considered before for the KRR models in Sec.~\ref{sec: dataset comparison}. The Adam optimizer~\cite{Kingma2014} 
is used to find the models parameters. In particular, at each step the gradient of the loss function is evaluated on a subset of the training set ("batch") of $N_{b} = 4$ configurations.
The results for the different datasets are reported in Fig.~\ref{fig: comparison no pressure MACE}, where we show the behavior of $\Delta_X$, defined in Eq.~\eqref{eq: Delta models}, for energy, forces, and pressure. 
They are consistent with those previously discussed in Sec.~\ref{sec: dataset comparison}, showing the effect of the SCE and the importance of correcting it regardless of the particular 
choice for the MLIP framework.

\subsection{Molecular dynamics simulation details}\label{appendix: MD details}

In this 
Appendix, we give further details on the MD simulations discussed in Sec.~\ref{sec: MD results}. 

At each step, the energy, forces, and pressure were calculated at the DFT level using the {\textsc{Quantum Espresso}} package in its GPU accelerated version~\cite{Giannozzi2009,Giannozzi2017,Giannozzi2020} with the PBE functional, and then summed with those predicted by the different MLIPs.  
For the DFT simulations, a $60$~Ry plane-wave cutoff with a projector augmented wave pseudopotential~\cite{PAWpseudo} was used together with a $4\times 4\times4$ Monkhorst-Pack $\mathbf{k}$-point grid.  For the MD simulations we used a time step of $0.25$~fs and a Langevin thermostat~\cite{Ricci2003,Attaccalite2008} with damping $\gamma = 0.13$~fs$^{-1}$.

\begin{figure}
    \centering
    \includegraphics[width=0.9\linewidth]{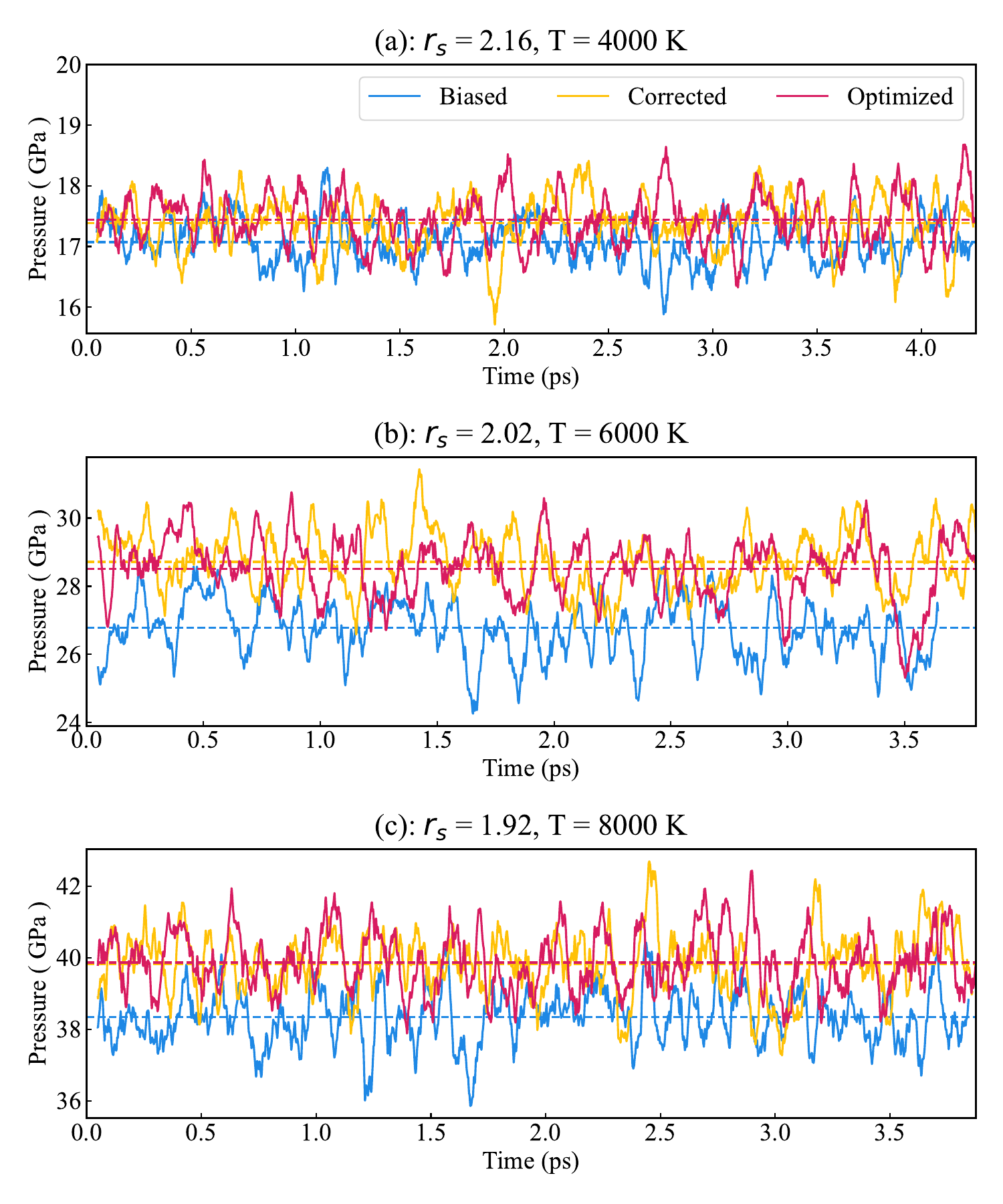}
    \caption{Pressure as a function of simulation time for a system of $128$ hydrogen atoms at different thermodynamic conditions: (a) $r_s = 2.16$ and temperature $T = 4000K$, (b) $r_s = 2.02$ and temperature $T = 6000K$, and (c) $r_s = 1.92$ and temperature $T = 8000K$. Each solid line corresponds to an MLIP trained on the different datasets while the dashed lines are the average values.}
    \label{fig: md result}
\end{figure}

After equilibration, we ran each simulation for about $4$~ps to obtain the equilibrium quantities. The behavior of pressure as a function of the simulation time for different densities and temperatures and for each MLIP is shown in Fig.~\ref{fig: md result}. 

\clearpage
\bibliography{Bibliography}
\end{document}